\documentstyle[aps,prc,floats,epsfig]{revtex}

\newcommand{\be}{\begin{equation}}
\newcommand{\ee}{\end{equation}}
\newcommand{\bea}{\begin{eqnarray}}
\newcommand{\eea}{\end{eqnarray}}
\newcommand{\bt}{\begin{tabular}}
\newcommand{\et}{\end{tabular}}
\newcommand{\ba}{\begin{array}}
\newcommand{\ea}{\end{array}}

\newcommand{\vp}{\varphi}

\begin{document}

\twocolumn[\hsize\textwidth\columnwidth\hsize\csname
@twocolumnfalse\endcsname

\small{\hfill{
\begin{tabular}{l}
DSF$-$23/2001
\end{tabular}}}


\title{Majorana solution of the Thomas-Fermi equation}
\author{Salvatore Esposito}

\address{Dipartimento di Scienze Fisiche, Universit\`{a} di Napoli
``Federico II'' and Istituto Nazionale di Fisica Nucleare, Sezione di
Napoli
\\ Complesso Universitario di Monte S. Angelo,  Via Cinthia, I-80126
Napoli, Italy
\\ E-mail: Salvatore.Esposito@na.infn.it }

\maketitle

\begin{abstract}
We report on an original method, due to Majorana, leading to a
semi-analytical series solution of the Thomas-Fermi equation, with
appropriate boundary conditions, in terms of only one quadrature. We also
deduce a general formula for such a solution which avoids numerical
integration, but is expressed in terms of the roots of a given polynomial
equation.
\end{abstract}

\vspace{1cm}

PACS numbers: 31.15.Bs, 31.15.-p, 02.70.-c

\vskip2pc]
\begin{figure}[t]
\epsfysize=6cm
\epsfxsize=8cm
\centerline{\epsffile{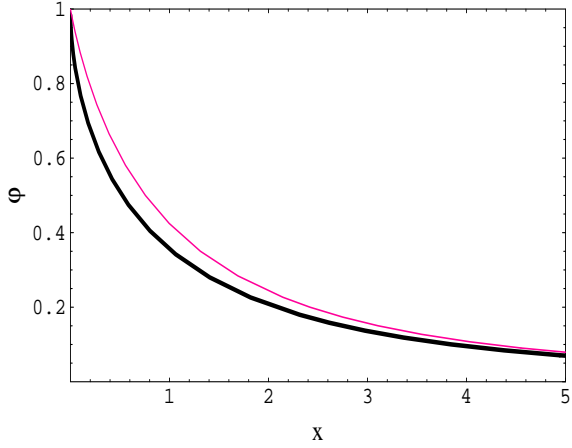}}
\caption{The Thomas-Fermi function $\vp(x)$. The thin (upper) line refers to the
exact (numerical) solution of Eq. (\ref{1}), while the thick (lower) one
corresponds to the parametric solution reported in Eqs. (\ref{29}).}
\label{f1}
\end{figure}

\noindent
In 1928 Majorana found an interesting semi-analytical solution of the
Thomas-Fermi equation \cite{thomfer} which, unfortunately, remained
unpublished and unknown until now (see \cite{volumetti}). We set forth here
a concise study of such a solution in view of its potential relevance in
atomic physics as well as in nuclear physics (as, for example, in some
questions related to nuclear matter in neutron stars \cite{nuastro}).
\\
The problem is to find the Thomas-Fermi function $\vp(x)$ obeying the
differential equation:
\begin{equation}\label{1}
  \vp^{\prime \prime} \; = \; \frac{\vp^{3/2}}{\sqrt{x}}
\end{equation}
with the boundary conditions:
\bea
\vp(0) &=& 1 \label{2} \\ \vp(\infty) &=& 0 \label{3} ~~~.
\eea
An exact particular solution of Eq. (\ref{1}) satisfying, however, only the
condition (\ref{3}), was discovered by Sommerfeld \cite{sommer}:
\begin{equation}\label{4}
  \vp \; = \; \frac{144}{x^3} ~~~.
\end{equation}
This can be regarded as an asymptotic expansion of the desired solution,
and Sommerfeld himself considered a ``correction'' (in some sense) to Eq.
(\ref{4}) in such a way to take into account the condition (\ref{3}).
However, this ``corrected'' approximate solution had a divergent first
derivative for $x=0$ \cite{sommer}.
\\
As will be clear below, Majorana solution can be considered as a
modification of Eq. (\ref{4}) as well, but the method followed by him is
extremely original and very different from the one used by Sommerfeld.
\\
Let us consider solutions of the Thomas-Fermi equation (\ref{1}) which are
expressed in parametric form:
\begin{equation}\label{5}
  \ba{rcl} \displaystyle x &=& \displaystyle x(t) \\ & & \\
  \displaystyle \vp &=& \displaystyle \vp(t) ~~~.
  \ea
\end{equation}
To be definite, throughout this paper we will use a prime ${}^\prime$ or a
dot $\dot{}$ to denote derivatives with respect to $x$ or $t$,
respectively. The strategy adopted by Majorana is to perform a double
change of variables:
\begin{equation}\label{6}
  x ~,~ \vp(x) ~~~~~ \longrightarrow ~~~~~ t ~,~ u(t)
\end{equation}
where the novel unknown function is $u(t)$. The relation connecting the two
sets of variables (assumed to be invertible) has a differential nature,
that is:
\begin{equation}\label{7}
  \ba{rcl}
  t &=& t(x,\vp) \\ & & \\ u &=& u(\vp, \vp^\prime) ~~~.
  \ea
\end{equation}
In such a way the second order differential equation (\ref{1}) for $\vp$ is
transformed into a first order equation for $u$. Note, however, that in
general Eqs. (\ref{7}) are implicit equations for $t$ and $u$, since $x$
and $\vp$ depend on them (one is looking for parametric solutions in terms
of the parameter $t$ and the unknown function $u$). For the specific case
of the Thomas-Fermi equation, Majorana introduced the following
transformation:
\bea
  t &=& 144^{-1/6} \, x^{1/2} \, \vp^{1/6}  \label{8} \\
  u &=& - \, \left( \frac{16}{3} \right)^{1/3} \, \vp^{-4/3} \, \vp^\prime
  \label{9} ~~~.
\eea
Observe that Eq. (\ref{8}) is reminiscent of the Sommerfeld solution, since
it can be cast into the form:
\begin{equation}\label{10}
  \vp \; = \; \frac{144}{x^3} \, t^6 ~~~.
\end{equation}
The differential equation for $u(t)$ is obtained by taking the
$t-$derivative of Eq. (\ref{9}):
\begin{equation}
  \frac{du}{dt} \; = \; - \, \left( \frac{16}{3} \right)^{1/3} \, \dot{x} \,
  \vp^{-4/3} \, \left[ - \, \frac{4}{3} \, \frac{\vp^{\prime 2}}{\vp} \, + \,
  \vp^{\prime \prime} \right]
\end{equation}
and inserting Eq. (\ref{1}):
\begin{equation}
  \frac{du}{dt} \; = \; - \, \left( \frac{16}{3} \right)^{1/3} \, \dot{x} \,
  \vp^{-4/3} \, \left[ - \, \frac{4}{3} \, \frac{\vp^{\prime 2}}{\vp} \, + \,
  \frac{\vp^{3/2}}{x^{1/2}} \right] ~~~.
\end{equation}
By using Eq. (\ref{8}) and Eq. (\ref{9}) to eliminate $x^{1/2}$ and
$\vp^{\prime 2}$, respectively, we obtain:
\begin{equation}\label{11}
  \frac{du}{dt} \; = \; \left( \frac{4}{9} \right)^{1/3} \, \frac{t u^2 -1}{t}
  \, \dot{x} \, \vp^{1/3} ~~~.
\end{equation}
We have now to express the quantity $\dot{x} \vp^{1/3}$ in terms of $t,u$.
From Eq. (\ref{8}),
\begin{equation}
  x \; = \; 144^{1/3} \, t^2 \, \vp^{1/3} ~~~,
\end{equation}
by taking the explicit $t-$derivative of both sides,
\begin{equation}
  \dot{x} \; = \; 144^{1/3} \, \left[ 2 t \, \vp^{-1/3} \, + \, t^2 \, \dot{x} \,
  \left( - \, \frac{1}{3} \, \vp^{-4/3} \, \vp^\prime \right) \right] ~~~,
\end{equation}
after some algebra we get:
\begin{equation}\label{11b}
  \dot{x} \, \vp^{1/3} \; = \; 144^{1/3} \, \frac{2t}{1 - t^2 u} ~~~.
\end{equation}
By inserting this result into Eq. (\ref{11}) we finally have the
differential equation for $u(t)$:
\begin{equation}\label{12}
  \frac{du}{dt} \; = \; 8 \, \frac{t u^2 - 1}{1 - t^2 u} ~~~.
\end{equation}
The condition (\ref{2}) implies, from Eqs. (\ref{8}),(\ref{9}), that $t=0$
for $x=0$ and:
\begin{equation}\label{13}
  u(0) \; = \; - \, \left( \frac{16}{3} \right)^{1/3} \, \vp^\prime_0
\end{equation}
where $\vp^\prime_0 = \vp^\prime(x=0)$. The initial condition to be
satisfied by $u(t)$ for the univocal solution of Eq. (\ref{12}) is obtained
from the boundary condition (\ref{3}) by inserting the Sommerfeld
asymptotic expansion (\ref{4}) into Eqs. (\ref{8}),(\ref{9}). For $x
\rightarrow
\infty$ we have $t=1$ and
\begin{equation}\label{14}
  u(1) \; = \; 1 ~~~.
\end{equation}
We then easily recognize that the branch of $u(t)$ giving the Thomas-Fermi
function (in parametric form) is the one between $t=0$ and $t=1$. In this
interval we look for the solution of Eq. (\ref{12}) by using a series
expansion in powers of the variable $\tau = 1-t$:
\begin{equation}\label{15}
  u \; = \; a_0 \, + \, a_1 \, \tau \, + \, a_2 \, \tau^2 \, + \, a_3 \,
  \tau^3 \, + \, \dots  ~~~.
\end{equation}
From the condition (\ref{14}) we immediately have:
\begin{equation}\label{16}
  a_0 \; = \; 1 ~~~.
\end{equation}
The other coefficients are obtained by an iterative formula coming from the
substitution of (\ref{15}) into Eq. (\ref{12}):
\begin{equation}\label{17}
  \sum_{k=0}^\infty \, \sum_{l=0}^\infty \, A(k,l) \, \tau^{k+l} \; = \; 0
\end{equation}
where:
\bea
A(k,l) &=& a_k \left[ (l+1) a_{l+1} - 2 (l+4) a_l + (l+7) a_{l-1} \right] +
\nonumber \\
&~& - \, (k+l+1) \delta_{l 0} a_{k+l+1} + 8 \delta_{k 0} \delta_{l 0}
\label{18}
\eea
(we define $a_{-1}=0$). Eq. (\ref{17}) can also be cast in the form
($k+l=m$, $l=n$):
\begin{equation}
  \sum_{m=0}^\infty \, \left( \sum_{n=0}^m \, A(m-n,n) \right) \, \tau^{m} \; = \; 0
\end{equation}
so that, for fixed $m$, the relation determining the series coefficients is
the following:
\bea
&~& \sum_{n=0}^m \, a_{m-n} \left[ (n+1) a_{n+1} - 2 (n+4) a_n +
\right. \nonumber
\\ &~& \left. + (n+7) (1 - \delta_{n0}) a_{n-1} \right] \; = \;
(m+1) a_{m+1} - 8 \delta_{m 0}  \label{19}
\eea
(we have explicitly used that $a_{-1}=0$), with $m=0,1,2,3, \dots$. The
equation (\ref{19}) for $m=0$:
\begin{equation}
  \left( a_0 - 1 \right) \, \left[ a_1 - 8 \left( a_0 + 1 \right) \right]
  \; = \; 0
\end{equation}
is identically satisfied due to Eq. (\ref{16}). For $m=1$ we have a second
degree algebraic equation for $a_1$:
\begin{equation}
  a_1^2 \, - \, 18 \, a_1 \, + \, 8 \; = \; 0
\end{equation}
of which we have to choose the smallest root (we are performing a
perturbative expansion):
\begin{equation}\label{20}
  a_1 \; = \; 9 - \sqrt{73} ~~~.
\end{equation}
The remaining coefficients are determined, using Eqs. (\ref{16}) and
(\ref{20}), by linear relations. In fact excluding the cases with $m=0,1$,
after some algebra Eq. (\ref{19}) can be written as:
\bea
a_m &=& \frac{1}{2(m+8)-(m+1) a_1} \left\{ \sum_{n=1}^{m-2} \, a_{m-n}
\left[ (n+1) a_{n+1} + \right. \right. \nonumber \\
&~& \left. - 2 (n+4) a_n + (n+7) a_{n-1} \right]  + a_{m-1} \left[ (m+7) +
\nonumber \right. \\
&~& \left. \left. - 2 (m+3) a_1 \right] + a_{m-2} \left[ (m+6) a_1
\right] \right\} ~~~.  \label{21}
\eea
Note that the sum in the RHS involves coefficient $a_i$ with indices $i
\leq m-1$, so that the relation in (\ref{21}) gives explicitly the value of
$a_m$ once the previous $m-1$ coefficients $a_{m-1},a_{m-2}, \dots
,a_2,a_1$ (and $a_0$) are known.
\\
The series expansion in (\ref{15}) is uniformly convergent in the interval
$[0,1]$ for $\tau$, since  the series made of the coefficients only,
$\sum_{n=0}^\infty \, a_n$, is convergent. In fact, by setting $\tau =1$
($t=0$) into (\ref{15}), from Eq. (\ref{13}) we have:
\begin{equation}\label{22}
  - \, \vp^\prime_0 \; = \; \left( \frac{3}{16} \right)^{1/3} \,
  \sum_{n=0}^\infty \, a_n ~~~,
\end{equation}
which shows that the sum of such a series is determined by the (finite)
value of $\vp^\prime_0$ ($\vp^\prime_0 \simeq 1.588$ and thus
$\sum_{n=0}^\infty
\, a_n \simeq 2.7746$). Note also that the coefficients $a_n$ are positive
definite and that the series in (\ref{22}) exhibits geometric convergence
with $a_n/a_{n-1} \sim 4/5$ for $n \rightarrow \infty$. The numeric values
of the first 20 coefficients are reported in Table I.
\begin{table}
\label{t1}
\caption{Numerical values for the first 20 coefficients for the series
expansion of the function $u(t)$ in Eq. (\ref{15}).}
\centering
\begin{tabular}{llcll}
\hline
$a_1$ & 0.455996 & ~~~ & $a_{11}$ & 0.0316498
\\
$a_2$ & 0.304455 & ~~~ & $a_{12}$ & 0.0252839
\\
$a_3$ & 0.222180 & ~~~ & $a_{13}$ & 0.0202322
\\
$a_4$ & 0.168213 & ~~~ & $a_{14}$ & 0.0162136
\\
$a_5$ & 0.129804 & ~~~ & $a_{15}$ & 0.0130101
\\
$a_6$ & 0.101300 & ~~~ & $a_{16}$ & 0.0104518
\\
$a_7$ & 0.0796352 & ~~~ & $a_{17}$ & 0.00840559
\\
$a_8$ & 0.0629230 & ~~~ & $a_{18}$ & 0.00676661
\\
$a_9$ & 0.0499053 & ~~~ & $a_{19}$ & 0.00545216
\\
$a_{10}$ & 0.0396962 & ~~~ & $a_{20}$ & 0.00439678
\\
\hline
\end{tabular}
\end{table}
Given the function $u(t)$ we have now to look for the parametric solution
$x=x(t)$, $\vp = \vp(t)$ of the Thomas-Fermi equation. To this end let us
put:
\begin{equation}\label{23}
  \vp(t) \; = \; \exp \left\{ \int_0^t \, w(t) \, dt \right\}
\end{equation}
where $w(t)$ is an auxiliary function to be determined in terms of $u(t)$,
and the condition (\ref{2}) (or $\vp(t=0)=1$) is automatically satisfied.
By inserting Eq. (\ref{23}) into Eq. (\ref{9}) and using (\ref{11b}) we
immediately find:
\begin{equation}\label{24}
  w \; = \; - \frac{6 u t}{1 - t^2 u} ~~~.
\end{equation}
Summing up, the parametric solution of Eq. (\ref{1}) with the boundary
conditions (\ref{2}), (\ref{3}) takes the form:
\begin{equation}\label{25}
  \ba{rcl}
x(t) &=& \displaystyle \sqrt[3]{144} \, t^2 \, e^{2 {\cal I}(t)} \\ & & \\
\vp (t) &=& \displaystyle e^{-6 {\cal I}(t)}
  \ea
\end{equation}
with:
\begin{equation}\label{26}
  {\cal I}(t) \; = \; \int_0^t \, \frac{u t}{1 - t^2 u} \, dt
\end{equation}
and $u(t)$ is given by the series expansion in (\ref{15}) with the
coefficients determined by (\ref{16}), (\ref{20}) and (\ref{21}). Eq.
(\ref{25}) represents the celebrated Majorana solution of the Thomas-Fermi
equation; it is given in terms of only one quadrature \footnote{Eq.
(\ref{25}) is, probably, the major result of this paper and was obtained by
Majorana. What follows is, instead, an original further elaboration of the
material presented above}.
\\
We have performed numerically the integration in (\ref{26}) stopping the
series expansion in (\ref{15}) at the terms with $n=10$ and $n=20$,
respectively, and compared the parametric solutions thus obtained from
(\ref{25}) with the exact (numerical) solution of the Thomas-Fermi
equation. We found that the two Majorana solutions approximate (for excess)
the exact solution with relative errors of the order of $0.1\%$ and
$0.01\%$, respectively.
\\
We can also obtain an approximate (by defect) analytic solution by
inserting the series expansion (\ref{15}) into the expression (\ref{26}):
\bea
  {\cal I}(t) &=& \int_{1-t}^1 \, \frac{u (1-\tau)}{1 - (1-\tau)^2 u} \,
  d \tau \; = \nonumber \\
  &=& \int_{1-t}^1 \, \frac{1 + b_1 \tau + b_2 \tau^2 + \dots}{c_1
  \tau + c_2 \tau^2 + \dots} \, d \tau  \label{27}
\eea
with:
\begin{equation}\label{28}
\ba{rcl}
b_n &=& a_n \, - \, a_{n-1} \\ & & \\ c_n &=& b_{n-1} \, - \, b_n
\ea
\end{equation}
for $n \geq 1$, while $b_0=1$ and $c_0=0$. Note that:
\begin{equation}
\ba{rcl}
b_n  \, < \, 0 &~~~~~~& {\mathrm for}~ n \, \geq \, 1 \\ & & \\ c_n  \, <
\, 0 &~~~~~~& {\mathrm for}~ n \, > \, 1
\ea
\end{equation}
(and $b_0 , c_1 >0$). If we neglect $O(\tau^2)$ terms in (\ref{27}), the
quantity ${\cal I}(t)$ is approximated by:
\begin{equation}
{\cal I}(t) \; = \; \frac{b_1}{c_1} \, t \, - \, \frac{1}{c_1} \, \log \,
(1-t)
\end{equation}
and, in terms of the original $a_n$ coefficients, the approximate
parametric solution of the Thomas-Fermi equation is:
\begin{equation}\label{29}
\ba{rcl}
x(t) &=& \displaystyle \sqrt[3]{144} \, t^2 \, (1-t)^{ -
\frac{2}{2-a_1}} \, e^{- 2 \frac{1-a_1}{2-a_1} t} \\ & & \\
\vp (t) &=& \displaystyle (1-t)^{\frac{6}{2-a_1}} \, e^{ 6
\frac{1-a_1}{2-a_1} t}  ~~~.
\ea
\end{equation}
In Fig. \ref{f1} we compare the above solution with the exact (numerical)
one.
\\
More in general, we can truncate the series in (\ref{27}) to a certain
power $\tau^k$ and thus the integrand function is approximated by a
rational function:
\begin{equation}
  F(\tau) \; \equiv \; \frac{1 + b_1 \tau + b_2 \tau^2 + \dots + b_k \tau^k}{c_1
  \tau + c_2 \tau^2 + \dots + c_k \tau_k} \; \equiv \; \frac{P(\tau)}{Q(\tau)} ~~~.
\end{equation}
Let us then assume that the roots $\tau_i$ ($i=1,2,\dots,k$) of the
polynomial in the denominator,
\begin{equation}
Q(\tau_i) \; = \; 0 ~~~,
\end{equation}
are known, so that we can decompose the function $F(t)$ in a sum of simple
rational functions \footnote{For simplicity we are also assuming that all
the zeros of $Q(\tau)$ are simple roots, as it is likely in the present
case. However, the generalization to the case in which multiple roots are
present is straightforward.}:
\begin{equation}
\frac{1}{Q(\tau)} \; = \; \frac{1}{c_k} \, \left( \frac{f_1}{\tau - \tau_1} \,
+ \, \frac{f_2}{\tau - \tau_2} \, + \, \dots \, + \, \frac{f_k}{\tau -
\tau_k} \right)
\end{equation}
and:
\bea
F(\tau) &=& \frac{1}{c_k} \, \left( 1 \, + \,  b_1 \tau \, + \, b_2
\tau^2 \, + \, \dots \, + \, b_k \tau^k \right) \, {\cdot} \nonumber \\
&~& {\cdot} \left( \frac{f_1}{\tau - \tau_1} \, + \,
\frac{f_2}{\tau - \tau_2} \, + \, \dots \, + \, \frac{f_k}{\tau -
\tau_k} \right) ~~~.  \label{30}
\eea
The expressions for the coefficients $f_i$ ($i=1,2,\dots,k$) in terms of
the roots $\tau_i$ are as follows:
\begin{equation}\label{31}
f_i \; = \; \prod_{\ba{c} l=1 \\ l \neq i \ea}^{\displaystyle k} \,
\frac{1}{\tau_i - \tau_l}  ~~~.
\end{equation}
By inserting the decomposition (\ref{30}) into Eq. (\ref{27}), the integral
${\cal I}(t)$ is thus given by a (double) sum whose generic element has the
following form:
\bea
&~& \int_{1-t}^1 \, \frac{\tau^n}{\tau - \tau_i} \, d \tau \; = \;
 - \tau_i^n \, \log \, \frac{1 - \tau_i - t}{1 - \tau_i} \, + \nonumber \\
&~& + \, \sum_{l=1}^n \, \left( \ba{c} n \\ l \ea \right) \,
\frac{\tau_i^{n-l}}{l} \, \left( (1-\tau_i)^l \, - \, (1 - \tau_i - t)^l \right) ~~~.
\eea
Then, in general, the parametric solution of Eq. (\ref{1}) can be formally
written as:
\begin{equation}\label{32}
  \ba{rcl}
x(t) &=& \displaystyle \sqrt[3]{144} \, t^2 \, \epsilon^2 (t) \\ & & \\
\vp (t) &=& \displaystyle \epsilon^{-6}(t)
  \ea
\end{equation}
where $\epsilon(t)=e^{{\cal I}(t)}$ is approximated by:
\bea
&~& \epsilon(t) \; = \;  \left[ \prod_{i=1}^k \, \left( \frac{1 - \tau_i -
t}{1 - \tau_i} \right)^{- \frac{1}{c_k} f_i \sum_{n=0}^k b_n
\tau_i^n} \right] \, {\cdot} \label{33} \\
&~& {\cdot} \, \left[ \prod_{i=1}^k \, \prod_{n=1}^k \, \prod_{l=1}^n \, e^{-
\frac{1}{c_k} f_i  \left( \ba{c} n \\ l \ea \right) \,
\frac{b_n \tau_i^{n-l}}{l} \, \left( (1 - \tau_i - t)^l  \, - \, (1-\tau_i)^l \right)}
\right] \nonumber
\eea
Obviously, using the method described above, the exact result is recovered
in the limit $k \rightarrow \infty$. This procedure can, however, be
employed for getting approximate but accurate solutions of the Thomas-Fermi
equation since, as it is clear from above, we have translated a numerical
integration problem (see Eq. (\ref{26})) into the one of a numerical search
for the roots of the polynomial $Q(\tau)$. Note also that we already know
one of such roots (namely, $\tau_1=0$) given the particular form of
$Q(\tau)$. This implies that, since the general solution of a fourth-degree
polynomial equation in terms of radicals is known, from (\ref{32}) and
(\ref{33}) we can get an analytic approximate solution by considering terms
in  the series in (\ref{27}) up to order $O(\tau^5)$, thus obtaining a
certainly much better approximation to Eq. (\ref{25}) than Eq. (\ref{29}).
We do not report here the explicit form of such a solution because of its
very long expression.
\\
Summarizing, in this paper we have reported on an original method, due to
Majorana, forwarding a semi-analytical solution of the Thomas-Fermi
equation (\ref{1}) with boundary conditions (\ref{2}), (\ref{3}). The
procedure applies as well to different boundary conditions, although the
constraint (\ref{2}) is always automatically satisfied. This corresponds to
physical situations present in atomic as well as in nuclear physics. We
have further studied the Majorana series solution thus obtaining a general
formula whose degree of approximation is limited by the one for searching
roots of a given polynomial rather than to the one for integrating a
rational function.
\\
The method used by Majorana for solving the Thomas-Fermi equation can be
generalized in order to study a large class of ordinary differential
equations, but this will be discussed elsewhere.



\acknowledgements
This paper takes its origin from the study of some handwritten notes by E.
Majorana, deposited at Domus Galileana in Pisa, and from enlightening
discussions with Prof. E. Recami and Dr. E. Majorana jr. My deep gratitude
to them as well as special thanks to Dr. C. Segnini of the Domus Galileana
are here expressed.

\end{document}